# Estimating wolf population size in France using non-invasive genetic sampling and spatial capture recapture models

Cyril Milleret[1,2,3*], Christophe Duchamp[2], Sarah Bauduin[4], Cecile Kaerle[5], Agathe Pirog[2,5], Guillaume Queney[5], Olivier Gimenez[1]

[1]CEFE, Univ Montpellier, CNRS, EPHE, IRD, Montpellier, France
[2]Service Conservation et Gestion des Espèces à Enjeux, Direction de la Recherche et de l'Appui Scientifique (DRAS), Office Français de la Biodiversité (OFB), Gap, France
[3]Faculty of Environmental Sciences and Natural Resource Management, Norwegian University of Life Sciences, Ås, Norway
[4] Service Conservation et Gestion des Espèces à Enjeux, Direction de la Recherche et de l'Appui Scientifique (DRAS), Office Français de la Biodiversité (OFB), Juvignac, France
[5] ANTAGENE, Animal Genomics Laboratory, La Tour de Salvagny (Lyon), France

*Corresponding author: cyril.milleret@gmail.com

**Acknowledgement** :

This study was made possible thanks to work of the professionals and volunteers from the wolf-lynx network operated by the DGPT at the French Biodiversity Agency. Partial funding came from the Research Council of Norway (NFR 286886; project WildMap). The computations were performed on resources provided by Norwegian University of Life Sciences's computing cluster 'Orion'.

**Declaration of competing interest**

The authors declare that they have no known competing financial interests or personal relationships that could have appeared to influence the work reported in this paper.

**Ethics statement**

None

**CRediT authorship contribution statement**

**Cyril Milleret** Writing – original draft, Methodology, Visualization, Validation, Formal analysis, Data curation, Conceptualization;**Christophe Duchamp**, Conceptualization, Methodology, Writing – review & editing, Conceptualization**; Sarah Bauduin**, Writing – review & editing, Validation, Data curation; **Cecile Kaerle,** Writing – review & editing, Validation, Data curation; **Agathe Pirog** Writing – review & editing, Validation, Data curation; **Guillaume Queney**, Writing – review & editing, Validation, Data curation; **Olivier Gimenez** Conceptualization, Methodology, Supervision, Writing – review & editing, Funding acquisition

**Abstract:**

Population size is a key metric for management and conservation. This is especially true for large carnivore populations for which management decisions are often based on population size estimates. In France, gray wolves (*Canis lupus)* have been monitored for more than two decades using non-invasive genetic sampling and capture-recapture models. Population size estimates directly inform the annual number of wolves that can be killed legally. It is therefore key to use appropriate methods to obtain robust population size estimates. To track the recent numerical and geographical expansion of the population, a substantial increase in sample collection was performed during the winter 2023/24 within the entire wolf distribution range in France. A total of 1964 samples were genotyped and assigned to 576 different individuals using microsatellites genetic markers. During the winter 2023/24, spatial capture-recapture models estimated the wolf population size in France to be likely between 920 and 1125 individuals (95% credible interval). Detection probability varied spatially and was positively influenced by snow cover and accessibility. Wolf density was strongly associated with the recent presence of the species, reflecting the ongoing recolonization process from the Alps. This work illustrates the usefulness of non-invasive genetic data and spatial capture-recapture for large-scale population assessment. It also lays the ground for future improvements in monitoring to fully exploit the potential of spatial capture-recapture models.

## 1. Introduction

Large carnivore populations have recolonized numerous parts of their former range (Chapron et al. 2014, Di Bernardi et al. 2025). As apex predators, they often play an important role in ecosystem functioning (Martin et al. 2020). At the same time, predation on domestic animals and game species are often the source of conflicts in human-dominated landscapes. This is particularly true for the gray wolf (*Canis lupus*) that fuels, at times, acrimonious public and political debates. Stakeholders involved in management and conservation are therefore in need of robust information, usually through the monitoring of the population, to determine management actions (Nichols and Williams 2006). From a management point of view, population size estimate is a key parameter. Estimates of population size are used to determine current and future population trends, and assess the conservation status of the species. It is also used to determine the number of individuals that can be legally removed(Andrén et al. 2020) . In this context, using methods that can provide robust estimates of population size are crucial.

The capture-recapture (CR) framework – which allows the estimation of detection probability from the multiple detection of identifiable individuals – is known as a robust method to account for the imperfect detection of individuals and obtain unbiased population size estimates (Williams et al. 2002, McCrea and Morgan 2014). More recently, spatial-capture recapture (SCR) models, a natural extension of CR models, have been developed to account explicitly for spatial heterogeneity in detectability linked with the location of individuals (Efford 2004, Royle and Young 2008, Borchers and Efford 2008). The estimation of population size with SCR models has been shown to be robust in many different situations (Theng et al. 2022), including when data are collected using non-invasive genetic sampling methods (Russell et al. 2012) or for species with a group living behavior (Bischof et al. 2020a). SCR models can be used to monitor many wildlife species in and is now used to monitor wolves in many countries (López-Bao et al. 2018, Bischof et al. 2020b, Marucco et al. 2023, da Costa et al. 2025, Iosif et al. 2025).

Wolves have recolonized France in the Alps from Italy in the early 1990's. Since then, the population has increased both numerically and in its distribution (Louvrier et al. 2018). The population is currently monitored by a multi-partner network of professionals and volunteers covering the entire country. Population size estimates are required to determine the annual number of wolves that can be legally shot (Arrêté du 23 octobre 2020, art. 1.). Monitoring and population size estimation rely on the non-invasive genetic sampling (NIGS) of individuals during winter (Cubaynes et al. 2010, Duchamp et al. 2012). Following the numerical and geographical expansion of the species, larger logistic and financial efforts were set to increase the number of detections and obtain more precise population size estimates.

Until the winter 2022/23, wolf population size estimates were obtained using non-spatial capture recapture models (Cubaynes et al. 2010). Following the new NIGS strategy of wolves in France and the development of SCR models that now allows large scale population size estimation (Bischof et al. 2020b, Marucco et al. 2023), we applied SCR models to estimate wolf population size in France. In addition, SCR models can identify factors involved in the spatial variation in detectability as well as determinants of spatial variation in density (Royle et al. 2014, Zhang et al. 2023, Moqanaki et al. 2023).

In this paper, we used wolf NIGS data and SCR models to estimate wolf population size in France during the winter 2023/24. Additionally, we aimed at quantifying sources of spatial and individual heterogeneity in detectability. We also expected wolf density to be correlated with spatial determinants linked with characteristics of the landscape but also to their recolonization history as it is often the case with large carnivore populations (Louvrier et al. 2018, Marucco et al. 2023, Moqanaki et al. 2023).

## 2. Material and methods

### 2.1. Data collection

The monitoring of wolves in France is coordinated by the French Biodiversity Agency and performed by the "wolf-lynx" network of more than 3500 specifically-trained members (Duchamp et al. 2012, Bauduin et al. 2023). Members of the network are mostly (70%) wildlife professionals and volunteers that collect biological samples on the presence of the species within the entire French administrative area. The main task of the network consists in collecting opportunistically non-invasive genetic samples (NIGS: scats, urine, hair, blood) during the winter (November 1-March 31). Until the winter 2003, no specific goals in terms of search-effort or collection of samples were given. From 2003 until 2022, the goals were to detect packs, collect samples within each known pack, and detect wolves as they expanded into new areas. Following the new national management plan updated in 2024, larger financial and logistic efforts were allocated to perform winter NIGS. In 2023/24, a 10*10km search grid was defined to stratify the sampling in two types of cells based on wolf presence during the previous winter (2022/23; Figure 1). In cells with sporadic wolf presence (n=155; no pack recorded) a goal of 4 collected samples was set. In cells with regular wolf presence (n=366; pack recorded), a goal of 8 collected samples was set. The goals were set to homogenize sample collection through the wolf population range while trading off the financial and logistical challenges that allowed the genetic analysis of 1860 samples within the grid at large scale. All samples collected outside of the grid during the winter were also genetically analyzed. During the NIGS, there was no direct recorded measure of search effort.

### 2.2. Genetic identification

Species and individual identification of each sample was performed through the sequencing of the mitochondrial control region and the genotyping of 23 (22 autosomal loci and a sex locus) microsatellite markers (Duchamp and Queney 2019, Pirog et al. 2025). Genotyping was systematically replicated four times on independent PCRs (multitube approach) to associate a quality index to each sample (Miquel et al. 2006). Only genotypes with an IQ>0.5 and presenting at least 12 loci without missing data were retained for individual identification (Pirog et al, 2025). Following the mismatch method, all possible pairs of genotypes were compared to identify individuals. Samples with identical genotypes were attributed to the same individual. To account for potential genotyping errors (such as allele dropouts and false alleles, which are common in NIGS approaches using degraded and low-quantity DNA, see (Waits and Paetkau 2005)), samples with three or fewer allelic differences (out of 44) were also assigned to the same individual. As frequently observed (Gagneux et al. 1997), false alleles were less frequent than allele dropouts (Pirog et al. 2025). In any case, microsatellite electropherograms of each replicate were checked to confirm individual assignation and to retain a consensus genotype for each individual. Details about the procedure are available in (Pirog et al. 2025).

### Spatial capture recapture model

We analyzed the NIGS data collected in the winter 2023/24 using a single-season spatial capture-recapture (SCR) model within a Bayesian framework (Royle et al. 2014). This model can cope with three challenges associated with population-level monitoring: 1) Because detection is imperfect, not all individuals present in the study area are detected; 2) Detection is also spatially heterogenous which means that the location of individuals influences their detection probability; 3) Individuals that reside primarily outside the surveyed area may still be detected within it. Without an explicit link between the population size parameter and the geographic area the population occupies, density cannot be estimated, and population size is ill-defined (Efford, 2004). The SCR model is composed of three submodels, one for population size, one for density, and one for detection of DNA samples.

#### 2.2.1. Population size

In this paper, we aimed at estimating overall wolf population size during winter. To estimate population size, we used the data augmentation approach (Royle and Dorazio 2012). This consists in augmenting the detection history dataset with $M$ individuals never detected. The model can therefore estimate the number of individuals that were not detected during the sampling. The state of an individual *i* is described by a Bernoulli state variable ($z_i$), which takes the value 1 if the individual is a true member of the population and 0 otherwise:

$$z_i \sim Bernoulli(\psi)$$

where $\psi$ is the probability that an individual in the detection history dataset is part of the population. With data augmentation, we need to choose $M$ sufficiently large so that $\psi$ does not include 1.

### 2.2.2. Density

In SCR models, the density sub-model describes the spatial configuration of individual activity centers (ACs; $s_i$). ACs are latent parameters describing the location of the center of activity of all individuals. We defined the habitat extent - area in which the individual activity centers can be located - as an area covering 100km around all confirmed wolf genotyped samples. The habitat extent was discretized in 10x10km habitat grid cells.

To account for the heterogeneous distribution of ACs, we used a Bernoulli point process (Zhang et al. 2023). Spatial variation in the distribution of individual activity centers was modeled using an intensity function:

$$\lambda(s) = e^{\beta X(s)}$$

where **X**($s$) is a vector of spatial covariate values evaluated at location $s$, and $\beta$ is a vector of associated regression coefficients. The intensity function $\lambda$ conditions the placement of AC within each cell of the habitat extent. In this formulation, no intercept is needed, as the number of activity centers is conditioned by data augmentation. Therefore, regression coefficients represent the relative effects of the different covariates on wolf density (Zhang et al. 2023)

We chose the covariates based on previous work performed by Marucco et al (2023) that estimated wolf population size in the alpine part of Italy. This population is connected with the French wolf population and shares comparable landscape characteristics, at least for the main range of the population located in the Alps. To model the spatial variation in AC density, we therefore considered the additive effect of 4 different covariates:

$$e^{\lambda(s)} = \beta_{Rec} * Rec_s + \beta_{For} * Forest_s + \beta_{Low} * Low_s + \beta_{Human} * Human_s$$

where $Rec$ represents the recent wolf presence (Appendix S1; Figure S1). Current wolf distribution is shaped by the recolonization history from the Italian Alps (Louvrier et al. 2018). To quantify the spatial heterogeneity of this process, we followed the approach proposed by (Marucco et al. 2023) and combined the 10x10km resolution grids of wolf presence produced by the Large Carnivore Initiative Europe (LCIE-IUCN) in 2014 (representing wolf distribution from 2007 to 2011) (Chapron et al., 2014) and 2018 (representing wolf distribution from 2012 to 2016) (IUCN Red List, Kaczensky et al., 2021). Wolf presence was classified as sporadic (1) or permanent (3) in both grids. We summed the two grids to create an index of the recent presence of wolves in each habitat grid cell (from 0, no wolf presence registered in the two periods, to 6, permanent presence during the two periods). Given the importance of recolonization history in explaining current carnivore density (Marucco et al. 2023, Moqanaki et al. 2023), we expected this covariate to be positively correlated with wolf density.

*Forest* is the percentage of forest cover (Appendix S1; Figure S1), which we calculated as the percentage of forest cover for each habitat grid cell by combining all forest categories (broadleaf, coniferous and mixed) in the Corine Land Cover (CLC) 2018, downloaded at the European Union's Copernicus Land Monitoring Service information (https://doi.org/10.2909/71c95a07-e296-44fc-b22b-415f42acfdf0). Subsequently, we calculated the percentage of area covered by the forest class for each cell of the habitat grid cells. We expected wolf density to be positively correlated with forest cover as it was found to be associated with higher probability of colonization (Louvrier et al. 2018) and density (Marucco et al. 2023).

*Low* is the percent low natural vegetation (Appendix S1; Figure S1). As for the forest cover, we used the 2018 CLC data. We aggregated the heathland, transitional shrubland and natural grassland categories, which fell under the broader classification to create the natural low vegetation layer, then calculated the percent cover for each habitat grid cell. We used this covariate as wolves are also present in areas with low vegetation cover such as in mountainous areas (Marucco et al. 2023).

Last, *Human* is the human population density (Appendix S1; Figure S1). Human density was obtained from the « Global Human Settlement Layer » (Schiavina et al. 2023) which represents 1x1km grid resolution of the human density in 2020. We then log-transformed human density. We expected human density to be negatively associated with wolf density, as human-dominated landscapes are usually associated with lower density (Marucco et al. 2023).

#### 2.2.3. Detection

The observation model allowed us to model the processes involved in the detection of non-invasive genetic samples. The searched area (area where detections were possible) was defined as all areas of the habitat extent overlapping within the French administrative border. However, we left a 20km buffer area around France (approximatively >3 times the scale parameter, see below for a definition), where individual ACs can be located but where no detection is possible (bordering countries, e.g., Italy, Switzerland or Spain). Although NIGS samples can be collected continuously in space, we discretized the search area in a 5*5km detector grid and use the center of these grids as detector locations (Russell et al. 2012, Milleret et al. 2018).

SCR models account for spatial and individual heterogeneity in detectability by modelling individual $i$ detection probability $p$ as a function of distance from the individual activity center $s$ at detector locations $j$. In general, SCR models assume that individual detection probability $p_{ij}$ decreases with distance from the individual activity center using a half-normal function such as:

$$p_{ij} = p_{0_{ij}} e^{-\frac{d_{ij}^2}{2\sigma^2}}$$

where $d$ represents the Euclidan distance from the individual activity center $s_i$ and the detector location $j$. The baseline detection probability $p_0$ is the detection probability at the activity center location and $\sigma$, the scale parameter representing the speed at which detection probability declines with distance to the activity center. The parameter $\sigma$ is directly linked with home range size (Royle et al. 2014).

To account for individual, spatial and temporal heterogeneity in detection probability, we included several additive linear effects on the baseline detection probability $p_0$ :

$$logit(p_{0_{ij}}) = \alpha_{Region[j]} + \beta_{Effort} * Effort_j + \beta_{Snow} * Snow_j + \beta_{Roads} * \log(Roads_j) + \beta_{Prevdets} * Prevdets_i + \beta_{Sex} * Sex_i$$

where $\alpha$ represents the intercept and the $\beta$'s are the regression coefficients between covariates and the detection probability.

We accounted for spatial variation in detection probability associated with the location of detectors using:

- A region-specific intercept $Region$. Intrinsic characteristics of a region, whether they are linked to the characteristics of the landscape or the sampling, may create spatial heterogeneity in detectability. We estimated an independent intercept for seven different administrative regions. Regions were defined based on the administrative borders of departments (Appendix S1; Figure S3). The departments with a low number of detections were aggregated with neighboring departments to allow the estimation of detection probability.
- Spatial variation in snow condition $Snow$. Snow on the ground facilitates the detection of individuals. We quantified snow cover during the winter of 2023/24 (November 1 -March 31) using the MODIS (0.1 degrees) data. We calculated the average snow cover within the winter using monthly MODIS layers in each detector cell (Appendix S1; Figure S2).
- Spatial variation in road density $Roads$. Accessibility, quantified by the presence of roads, can increase detection probability. We used Open Street Map data on roads to quantify the number of kilometers per $km^2$ in each detector cell. We log-transformed the variable to give less importance to the area with high road density (Appendix S1; Figure S2).
- Spatial variation in effort $Effort$. Search effort is assumed to be correlated with detection probability. Wolf monitoring relies on trained professionals and volunteers, but there is no direct record of effort (e.g., number of kilometers traveled) during the field activity. To obtain a proxy of the *actual* sampling effort, we used an internal OFB database in which field activities related to the wolf-lynx network activities were recorded. We calculated the number of visits within each municipality. This variable does not represent the total effort though, as 54% of indices were collected by non-OFB members in 2023-2024 (Appendix S1; Figure S2).
- During preliminary analyses, we also tested a spatial covariate representing variation in the number of active members of the wolf-lynx network (Bauduin et al. 2023). However, this covariate showed no evidence of positive correlation, likely because it reflects *potential* rather than *actual* monitoring effort during winter 2023/24 (see supplementary S2).

To account for individual variation in detection probability, we used:

- The genetic detection of individuals for ≥2 winters over the three previous winters (2020/21-2022/23) $Prevdets$. The hypothesis is that an individual previously detected may have a higher detection probability because the approximate location of its home range is known Also, an individual previously detected may have already reached the stage adult, and we know that adult scent-marking individuals have a higher detection probability compared to younger individuals (Åkesson et al. 2022). This variable is also partially latent, as undetected individuals in the winter 2023/24 could have also been detected in ≥2 winters during the three previous winters (Milleret et al. 2021):

$$Prevdets_i \sim Bernoulli(probDet)$$

  where $probDet$ represents the probability that an individual in the population was detected in ≥2 winters during the three previous winters.

- The sex $Sex$ of the individual as a binary variable ($Sex$= 0 if female, and $Sex$ = 1 if male). Because the sex of the individual is not known for augmented individuals as well as for a few detected individuals, sex is a partially latent variable such as:

$$Sex_i \sim Bernoulli(probMale)$$

where $probMale$ represents the probability that an individual in the population is a male.

### 2.3. Model fitting

We standardized all covariates and we fitted the Bayesian SCR model using Markov Chain Monte Carlo using NIMBLE 1.3.0 (de Valpine et al. 2017, 2024) and nimbleSCR 0.2.0 (Bischof et al. 2020c, Turek et al. 2020) in R 4.3.0 (R Core Team 2023). We used computational techniques described in (Milleret et al. 2019, Turek et al. 2021) to speed up the computation of the model. We ran 4 chains of 37500 iterations with a burn-in of 7500 iterations. We assessed the convergence of the model using Gelman-Rubin diagnostic Rhat <1.1 (Gelman and Rubin 1992) as well as by visually inspecting the mixing of the parameters. To obtain population size estimates, we only considered individuals alive ($z_i = 1$) with their activity centers within the French administrative borders, therefore excluding individuals with their activity in areas where detection was not possible.

## 3. Results

A total of 1964 samples collected during the winter 2023/24 were analyzed genetically, of which 952 samples were successfully genotyped with sufficient quality to assign a wolf-ID (Figure 1). This corresponded to 467, 469 and 16 detections of female, male, and undetermined sex individuals, respectively. Overall, 576 different individuals ($n_{females}$=276, $n_{males}$=284, $n_{unknown}$=16) were detected, which led to an average of 1.65 detection per individual detected.

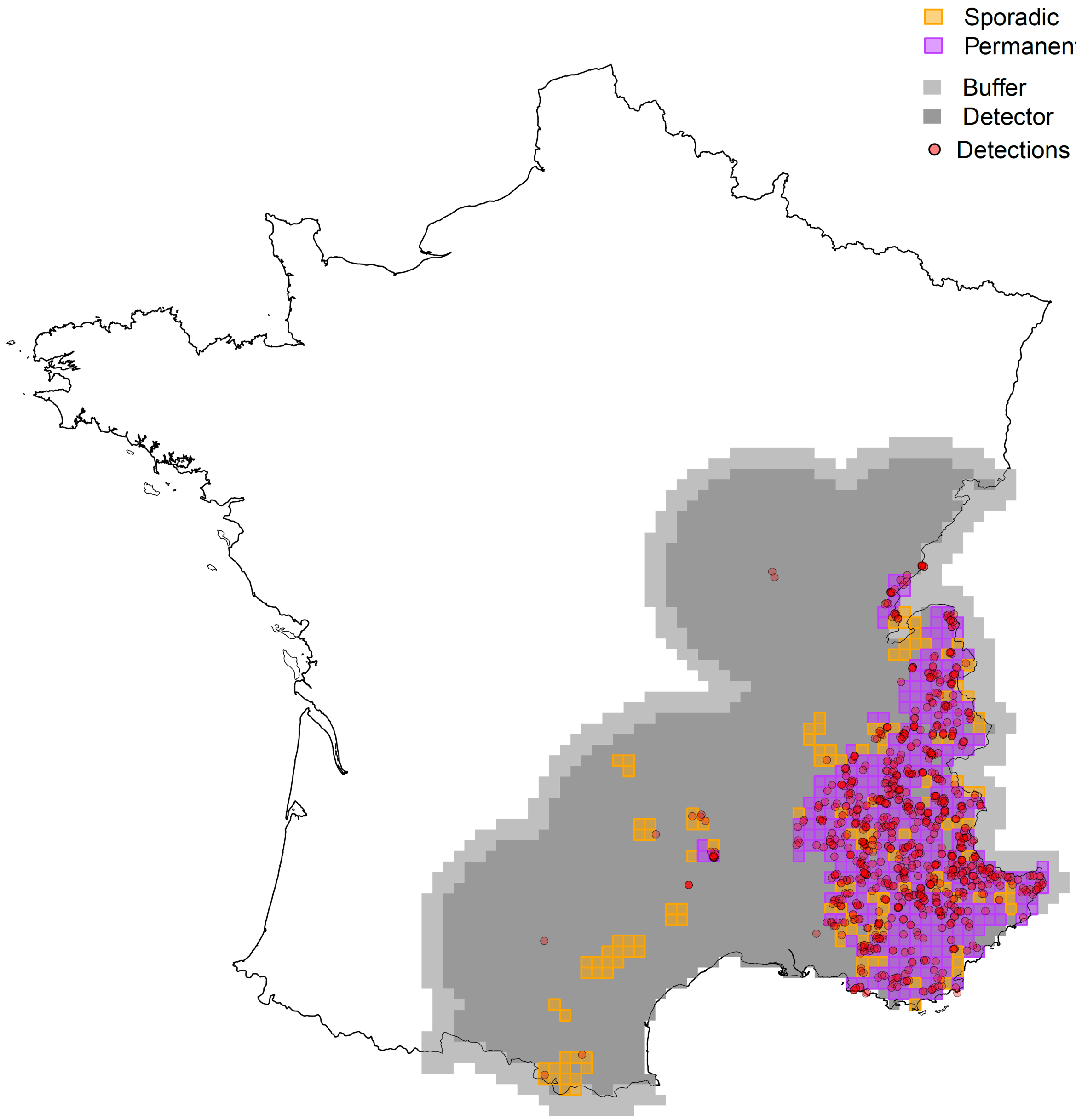

**Figure 1.** Map of the habitat range (light and dark grey area) considered in the SCR analysis covering an area of 100km around all genetically confirmed wolf detections (red dots) collected during the winter 2023/24. The 10x10km grid used to guide searches during sampling (sporadic = orange) and permanent (purple) where the goal was to collect 4 and 8 samples within each cell, respectively. The entire dark grey area was covered with a 5*5km detector grid for SCR analysis.

Wolf density was correlated positively with recent presence ($\beta_{Rec} = 0.89; 95\%\ ICr = 0.78; 1.01$), forest ($\beta_{For} = 0.40\ [0.28; 0.52]$), and cover of low natural vegetation ($\beta_{Low} = 0.14\ [0.02; 0.26]$). Human population density was negatively correlated with wolf density ($\beta_{Human} = -0.13\ ; [-0.29; 0.02]$). Detection probability varied spatially, according to the different types of regions (Appendix S1; Figure S3), and was correlated positively with the presence of snow ($\beta_{Snow} = 0.18; [0.10; 0.26]$) and density of roads ($\beta_{Roads} = 0.21\ ; [0.09; 0.35]$). Individuals detected in ≥ 2 winters (within the last three winters) had a higher chance of being detected $\beta_{PrevDet} = 0.40; [0.15; 0.65]$). Detection probability of males had the tendency to be lower than the one from the females ($\beta_{Sex} = -0.12$ ; $[-0.34; 0.09]$). However, there was no marked signal between detection probability and the effort covariate ($\beta_{Effort} = 0.03$ ; $[-0.04; 0.09]$). Finally, the scale parameter $\sigma$ of the detection function was estimated at $4.08km$ $[3.87km; 4.32km]$.

The size of the wolf population was estimated to be likely (95% credible interval) between 920 and 1125 individuals in 2023/24 (mean=1013). Given that 576 individuals were detected, this means that the individual detection probability was estimated to be 56.9% [51-62%] in 2023/24.

## 4. Discussion

In this study, we applied spatial capture-recapture models to estimate wolf population size in France from non-invasive genetic sampling data. SCR models exploit the spatial information contained in individual detections to account for individual and spatial variation in detectability, as well as for the inhomogeneous distribution of individuals within the landscape (Efford 2004, Royle and Young 2008, Borchers and Efford 2008, Royle et al. 2014). As expected, we found a strong association between current wolf density and spatial determinants linked with the recolonization history of the species and landscape characteristics. Our study is an important step towards obtaining robust and precise wolf population size estimates, which is key for wolf management in France.

We found that detection probability was positively associated with the presence of snow and with accessibility, represented by the density of roads. Both factors are known to help detection of wolf tracks and DNA material (Bischof et al. 2020b). Detectability also varied substantially among regions (Appendix S1; Figure S3). Our aim in including these regions was to capture large-scale differences in detectability related to sampling, landscape characteristics, or species biology (e.g. home range size) that were not explained by the spatial descriptors of the landscape we used. However, the variables included did not capture fine-scale variation in detectability, which could be linked, for example, to local snow conditions. The ability to account for both individual and spatial heterogeneity in detectability is a major strength of SCR models for obtaining robust population size estimates. This is particularly important when producing large-scale estimates of population size, where monitoring schemes, landscape characteristics, and even the behavior of monitored individuals are likely to be spatially heterogenous (Bischof et al. 2020b, Marucco et al. 2023).

As is often the case, large carnivore distribution reflects recolonization history and/or management policies (Marucco et al. 2023, Moqanaki et al. 2023). Here, we found that wolf density was correlated with areas of recent presence. In France, this recent presence is linked to the recolonization event that began in the 1990s, when wolves from the Italian population expanded into the French Alps. Since then, the wolf population steadily increased in both range and size (Louvrier et al. 2018). Wolves now occupy almost the entire alpine region, which constitutes the core of the population size at the national scale. Other covariates indicated that wolf density was lower in areas of high human density, but higher in areas with forest cover and low natural vegetation typical of Alpine landscapes. This pattern suggests a general avoidance of human-dominated areas, consistent with existing knowledge of the French population (Louvrier et al. 2018) and neighboring Italian population (Marucco et al. 2023).

The habitat extent was defined as the area within 100km of all confirmed individual detections obtained from NIGS during the winter 2023/24. This extent was defined arbitrarily, with the aim of avoiding the inclusion of large areas lacking individual detections. Estimating population density in areas with no or very few detections is challenging, as the model must rely solely on the prediction of density based on the spatial covariates used and on predicted detection probability. This issue is especially relevant in our case because no direct measure of search effort was available, and because landscape characteristics differ substantially between the area occupied by the core of the population and the current colonization front. Given the large dispersal abilities of wolves (Ražen et al. 2016, Konec et al. 2024), individuals are occasionally observed outside the defined habitat. However, these likely represent very few individuals, and for estimating total population size, it is cost-effective to focus both field and statistical efforts within the main population range, where most of the population is located. We identify the definition of habitat extent as an important aspect requiring further refinement in future wolf population size assessments (Milleret et al. 2025).

Large-scale monitoring programs are logistically and financially costly to initiate and sustain, especially for carnivores with elusive behavior and large home ranges. Involving multiple stakeholders is therefore a common strategy for monitoring wolf populations at large spatial scales (Ražen et al. 2020, Marucco et al. 2023). The French wolf-lynx network similarly includes a wide range of contributors, from professionals in various state agencies to volunteers such as hunters, naturalists and shepherds. This collaborative approach offers many advantages but also presents logistical challenges. In our case, it was not possible to record search effort, forcing us to rely on a proxy, which showed no evidence of a correlation with detection probability. Search effort by members of other state agencies or volunteers was not accounted for with this proxy, which likely explains the lack of correlation between this covariate and detection probability. Nevertheless, SCR population size estimates are generally robust to unmodelled variation in search effort, provided there are no large spatial gaps in sampling (Moqanaki et al. 2021). In this respect, the winter 2023/24 sampling strategy ensured coverage of all defined grids and avoided major spatial gaps. We also accounted for large-scale heterogeneity in detectability related to accessibility, snow conditions, and administrative regions. While methods have been proposed to account for latent heterogeneity in detectability, their application remains computationally prohibitive at large spatial scales (Dey et al. 2023a). Recording fine-scale spatial variation in search effort (e.g., number of kilometers searched, number of visits) would not only improve population size estimation (Moqanaki et al. 2021) but also help optimize the efficiency of the current monitoring program (Boiani et al. 2024). Until such data becomes available, we refrain from using SCR models to produce density maps or regional population size estimates (Bischof et al. 2020b).

By setting a target number of samples to be collected, the 2023/2024 monitoring strategy ensured sampling across the entire area of wolf presence, with detection probability > 0 in all occupied areas. This sampling strategy allowed us to use spatial capture-recapture methods instead of non-spatial methods (Cubaynes et al. 2010). With SCR models, both spatial and individual heterogeneity in detectability can be considered which makes SCR robust to a wide range of conditions (Bischof et al. 2020a, Dey et al. 2022, Theng et al. 2022, Efford 2025), even when detailed search effort is unavailable (Moqanaki et al. 2021). By leveraging spatial information from individual detections, SCR models also provide opportunities for estimating region-specific population sizes (Bischof et al. 2020b), spatial survival (Milleret et al. 2023), and movement (Efford and Schofield 2022, Kervellec et al. 2023). However, estimating the population size of an elusive, wide-ranging species at a large scale remains economically costly, and logistically and statistically challenging (Bischof et al. 2020b). Given the current spatial expansion of wolves in France, maintaining such a sampling strategy might be become logistically and financially costly in the future. Consequently, improving the cost-efficiency of monitoring programs (Boiani et al. 2024), integrating multiple sources of data (Gervasi et al. 2024) or considering alternative metrics that do not require explicit population size estimation (Bauduin et al. 2025) might become necessary for effective wolf population monitoring.

## Management implication

In France, wolf population size estimates obtained from capture-recapture models are used to determine the maximum number of individuals that can be legally shot (Arrêté du 23 octobre 2020, art.1). Using methods that have been shown to provide robust estimates is therefore essential for the management and conservation of the species. In this paper, our objective was not to compare different approaches — that all have their own advantages and limitations — but rather to demonstrate the application of spatial capture–recapture (SCR) models to estimate wolf population size in France. Although population size estimates from SCR models are considered as robust and generally more precise than non-spatial capture-recapture models (Blanc et al. 2013), biases in population size estimates linked with the behavior of wolves (Bischof et al. 2020a), genetic errors (Kodi et al. 2024), unaccounted individuals (Cubaynes et al. 2010) or spatial heterogeneity in detectability (Moqanaki et al. 2021) can occur. In this context, 1) reliable goodness-of-fit tests specifically developed for SCR models for assessing model performance and identifying potential sources of bias (Dey et al. 2022, Choo et al. 2024, Efford 2025), and 2) methods allowing to account for unknown sources of spatial and individual heterogeneity in detectability with sparse SCR datasets (Dey et al. 2023b) are needed.

The maximum number of wolves that can be legally culled is currently set to 19% of the annual mean population size estimate. Regardless of the method used to estimate population size, such estimates are subject to uncertainty and potential bias. However, current management decisions are based on a single point estimate and do not allow uncertainty to be formally incorporated into decision-making. Fully incorporating uncertainty would also support an adaptive management framework, in which the consequences of different culling intensity could be updated annually while taking into account management objectives, regulations, and new knowledge (Marescot et al. 2013, Andrén et al. 2020). In this context, we emphasize the importance of reporting, communicating, and using uncertainty measures, such as 95% credible intervals.

## Data availability:

Data and code to reproduce the analysis and the results are available on zenodo https://doi.org/10.5281/zenodo.18258795

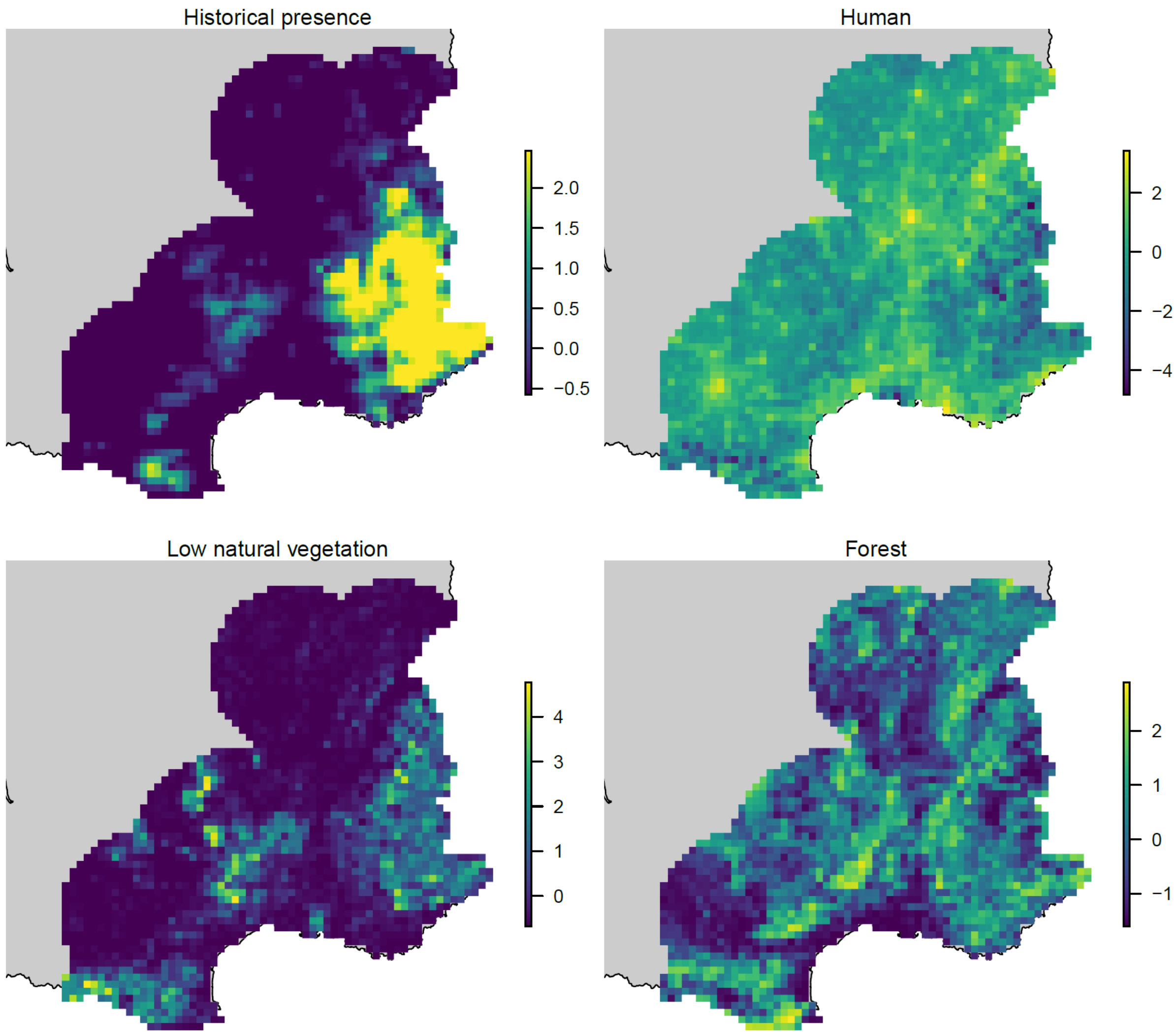

**Figure S1**. Habitat covariates used to model spatial variation in wolf density from non-invasive genetic sampling data collected during the winter 2023/24 and using SCR models. All covariate values were scaled prior to the analysis. A full description of the covariates is available in the main text.

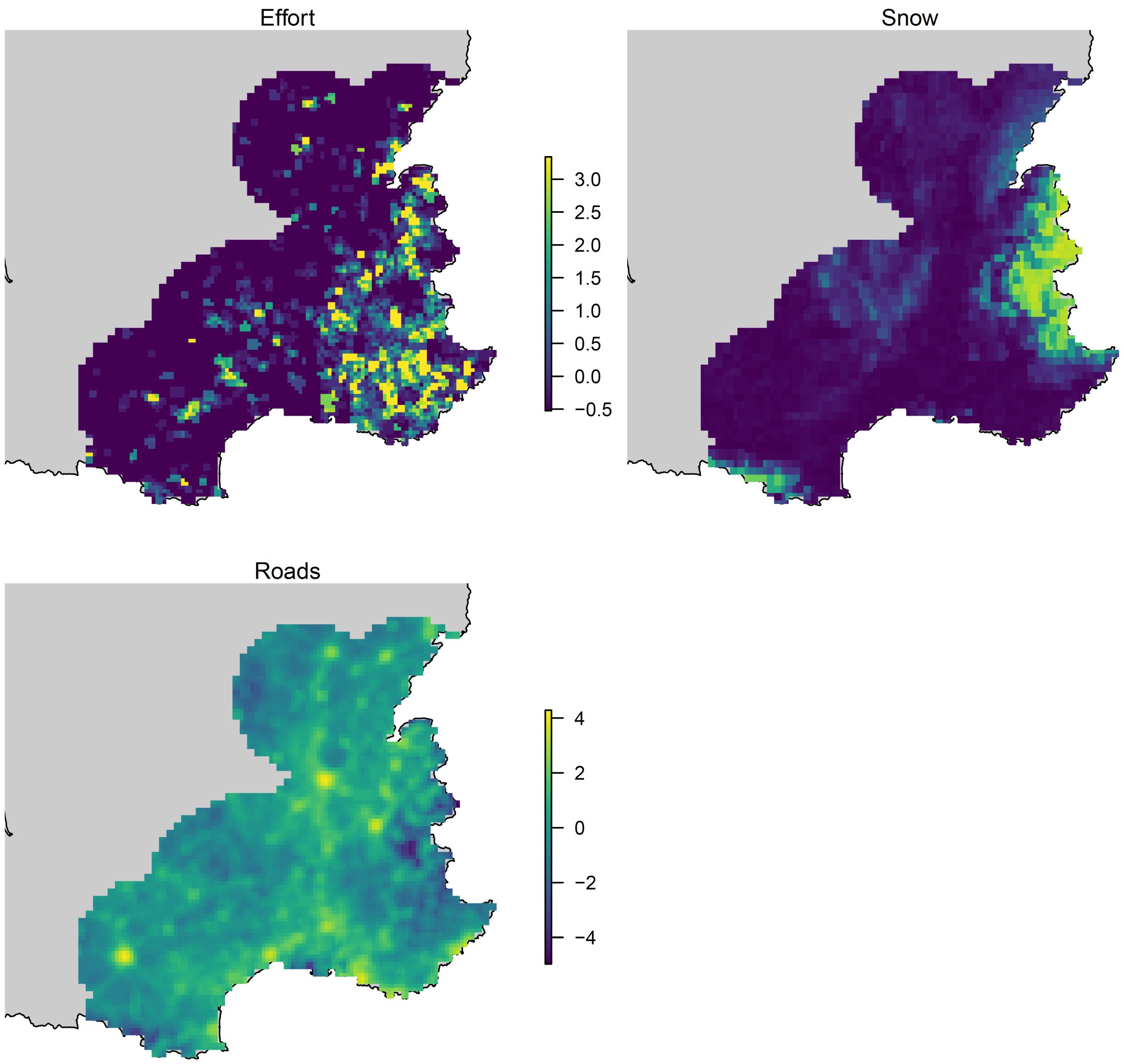

**Figure S2**. Detector covariates used to model spatial variation in the baseline detection probability ($p_0$) using non-invasive genetic sampling data collected during the winter 2023/24 and using SCR models. All covariate values were scaled prior to the analysis. A full description of the covariate is available in the main text. The baseline detection probability quantifies the probability to detect an individual right at its activity center location.

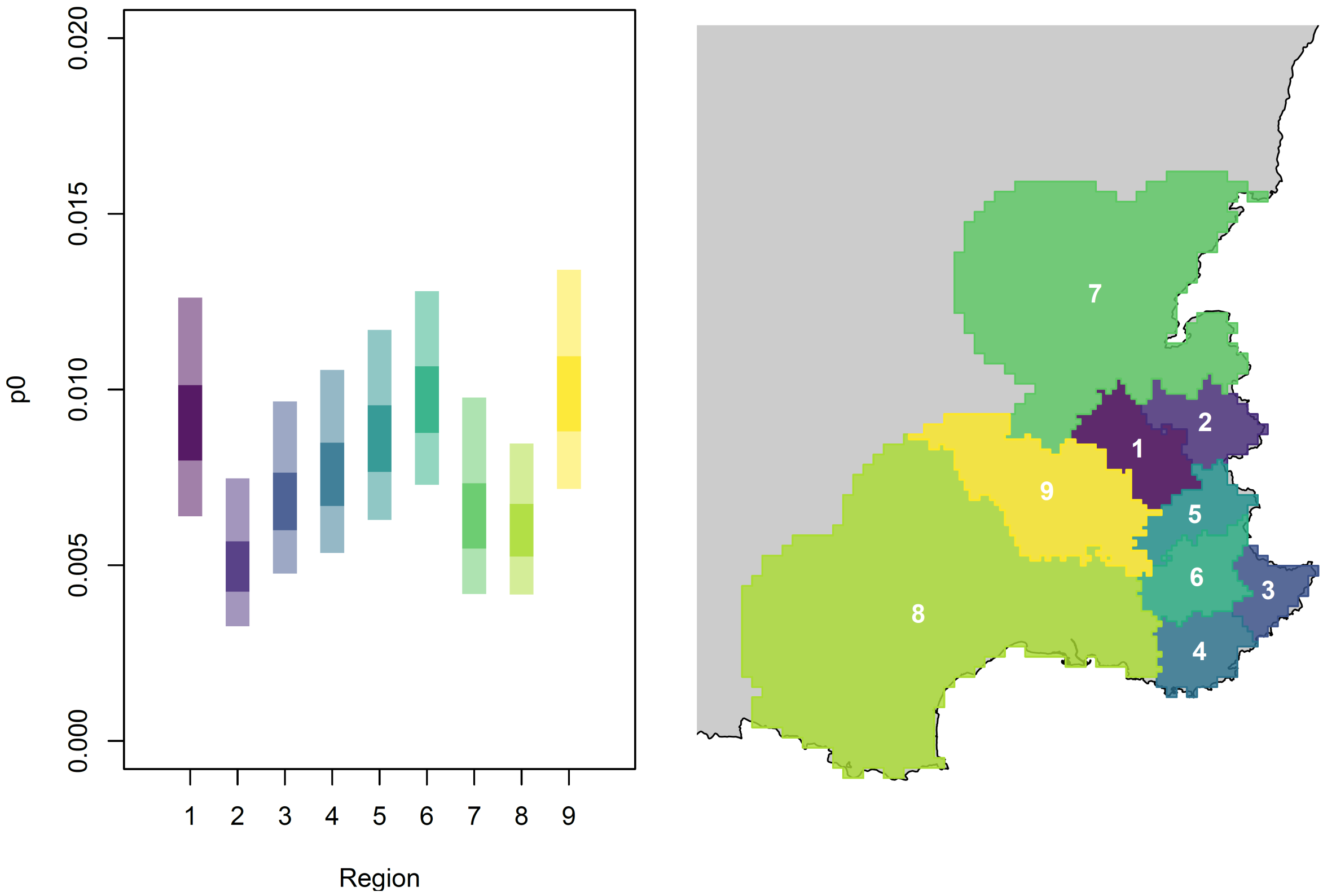

**Figure S3**. Region-specific estimates of the baseline  detection probability (p0) obtained from the spatial capture recapture model on the non-invasive genetic sampling data of wolves collected during the winter 2023/2024. Regions followed the administrative boundaries of the department. However, we merged administrative regions for the region 7, 8, and 9 to cope with the low number of NGS samples. The baseline detection probability quantifies the probability to detect an individual right at its activity center location.

## Appendix S2

## Alternative representation of search effort

Because direct record of search effort was not available, we looked for alternative proxies. We used a covariate that quantified the density of registered members of the French wolf-lynx network (Louvrier et al. 2018, Bauduin et al. 2023). The covariate was constructed to account for the area of prospection of the different network members using k-means clustering method (Wang and Song 2011, Bauduin et al. 2023). The clustering method accounted for variation in the extent of the area searched by the diversity of network members related to the nature of their job position and/or affiliation (Bauduin et al. 2023). This search effort covariate does not quantify realized search effort, but rather a “maximum potential” through the availability of registered network members. We ran the same SCR model described in the main text but replaced the search effort covariate by the covariate described above (Figure S4). We found that the density of active network members was negatively correlated with detection probability (mean=-0.16; 95% CrI= -0.25; -0.06), but its inclusion did not affect population size estimates (mean= 1009; 95% CrI= 920-1108). Search effort would be expected to correlate positively with detection probability, but we do not know the reason for the observed negative correlation. There are a few hypotheses for this unexpected result. First, this variable quantifies the potential effort but we do not have record of the actual active members compared to those registered who did not prospect in a given winter. There is a larger number of registered members in non-prospected areas compared to a fewer registered members in actually highly prospected areas. This could have inverted the expected relationship. For example, as in 2020, only 1687 members had recorded at least one sample among almost 4000 members registered in the network. Second, given the high goal in terms of number of samples to collect for the winter 2023/24, the additional search effort to reach this goal was covered by French Biodiversity Agency members. However, they only represent a quarter of all the registered members of the network. We expect realized search-effort to be significantly different, especially at the local scale from potential effort quantified with this variable.

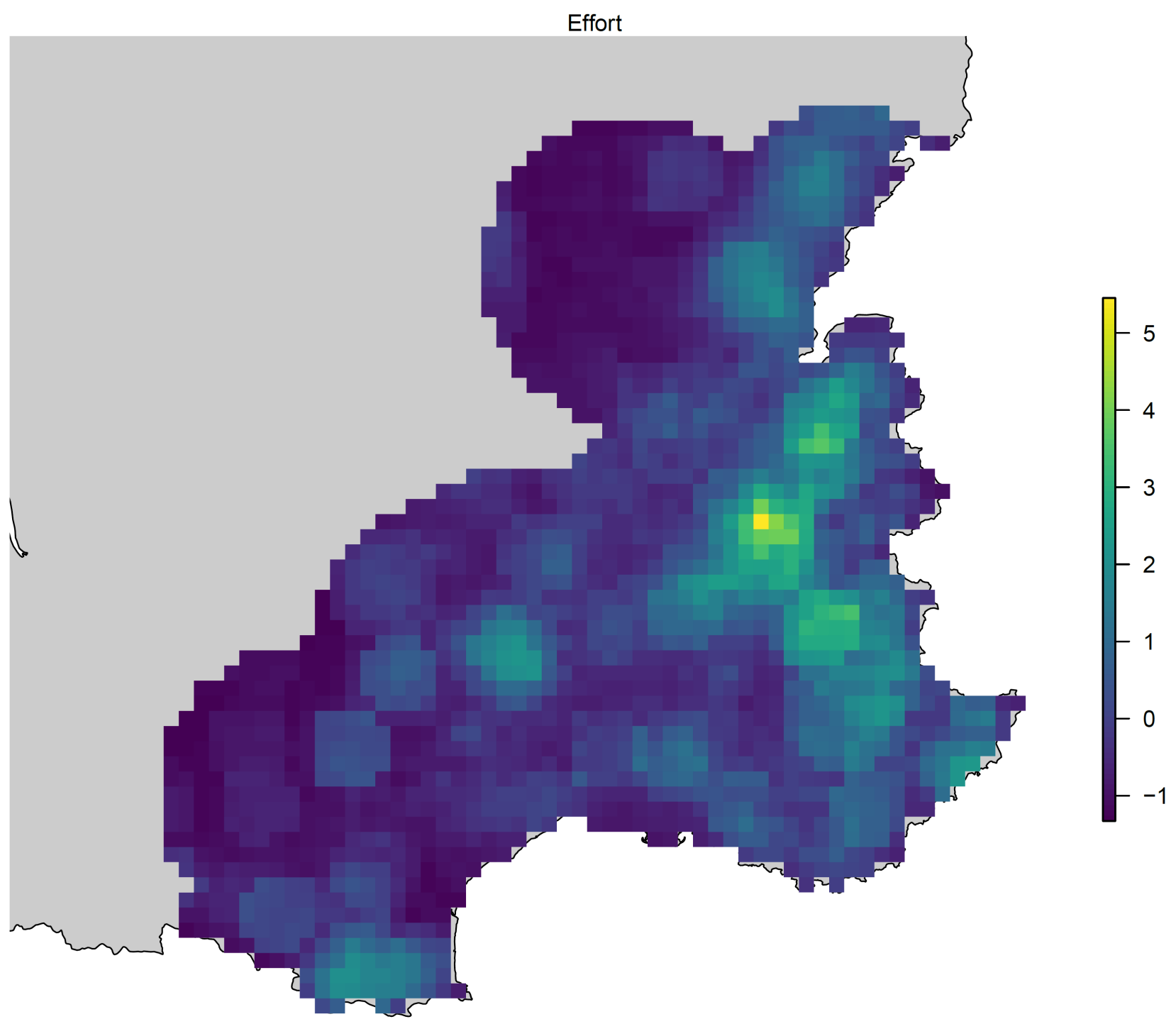

**Figure S4**. Detector covariate representing an alternative description of search effort that represents spatial variation in the availability of registered members of the wolf-lynx network (Bauduin et al. 2023). This covariate was used to model spatial variation in detection probability ($p_0$) using non-invasive genetic sampling data collected during the winter 2023/24 and using the SCR model described in the main text. The covariate values were scaled prior to the analysis.